\documentclass[twoside,leqno,twocolumn]{article}

\usepackage[letterpaper]{geometry}

\usepackage{ltexpprt}
\usepackage{graphicx}
\usepackage{xcolor}
\newlength{\wdth}

\begin{document}

\title{\Large Robustness of ML-Enhanced IDS to Stealthy Adversaries}
\author{Vance Wong\thanks{Laboratory for Advanced Cybersecurity Research, National Security Agency} \and John Emanuello \thanks{Laboratory for Advanced Cybersecurity Research, National Security Agency}}


\maketitle

\begin{abstract}\small
  Intrusion Detection Systems (IDS) enhanced with Machine Learning (ML) have demonstrated the capacity to efficiently build a prototype of ``normal'' cyber behaviors in order to detect cyber threats' activity with greater accuracy than traditional rule-based IDS. Because these are largely black boxes, their acceptance requires proof of robustness to stealthy adversaries. Since it is impossible to build a baseline from activity completely clean of that of malicious cyber actors (outside of controlled experiments), the training data for deployed models will be poisoned with examples of activity that analysts would want to be alerted about. We train an autoencoder-based anomaly detection system on network activity with various proportions of malicious activity mixed in and demonstrate that they are robust to this sort of poisoning.
\end{abstract}

\section{Introduction}
\indent \indent Malicious cyber activity is ubiquitous and the societal and economic impacts of sophisticated cyber attacks are immense\cite{cybercrime,WHCEA}. Given the shortage of cybersecurity professionals, automated tools are required \cite{IDS1, Lazarevic2005}. However, these so-called intrusion detection systems (IDS), which are largely rules-based systems, often fail to detect novel cyber attacks and produce more alerts than analysts can address.

The success of machine learning architectures in cybersecurity has created a new generation of IDS which show promise to be critical components of automated defense systems\cite{Heard,UNSW3, Lunt,verma}. Unlike prior iterations of these utilities, which relied on frail rules incapable of adapting to changing cyber threats, these can learn complex behaviors indicative of normal user activity and detect those which deviate from the baseline. This has the potential to enable detection of novel malicious activity, including zero-day attacks which the model has never seen\cite{UNSW4}. Such techniques, including autoencoder (AE) anomaly detectors, have been shown to be superior to rules-based IDS in detecting malicious activity on ground-truth data\footnote{The use of autoencoders for anomaly detection were previously shown to be effective in image forensics \cite{Cozzolino} and in manufacturing \cite{Liao}.}\cite{Nguyen}.

The ability of sophisticated adversaries to cover their tracks and make their activities look normal can potentially degrade the performance of these anomaly detection schemes if they are not carefully controlled. Because these are black-box techniques it can be difficult to know if they are learning the right characteristics of normal activity.

In this work, we introduce a combined feature engineering technique and AE-based anomaly detection scheme and demonstrate its robustness to stealthy adversaries. The features we engineer are derived from a technique traditionally applied to natural language processing, and, by construction, are directly relevant to the task of detecting anomalous behaviors using an AE. We demonstrate that our pipeline is robust to stealthy adversaries, by training on \emph{acceptable levels} of malicious behaviors and demonstrating that performance is maintained. We believe this sort of robustness demonstration to be the first of its kind. In a departure from some research, we conduct experiments on publicly available data, which enables our technique to serve as a baseline for others to improve upon.

Our paper is organized as follows: in Section \ref{sec:arch} we discuss an architecture for a neural IDS which is similar to others seen in the literature, introduce an open-source data set, and lay out the experiments we conduct; in Section \ref{sec:results} we present results and discuss the impact thereof in Section \ref{sec:discussion}.

\section{Proposed Model and Experimental Paradigm}\label{sec:arch}
\indent \indent In this work we make use of an autoencoder-based anomaly detection scheme. Autoencoders (AE) are neural networks, trained in an unsupervised fashion, wherein the target function to be learned is an approximate identity on the input space\cite{Goodfellow}. These networks are usually decomposed into non-linear encoder and decoder functions such that the image of the encoder $h$ is of much smaller dimension than the input space:
$$f(x)=g\circ h(x),$$
In essence, the AE is a lossy compression-decompression scheme, with the property that reconstruction will be poor on data drawn from a different distribution than the one that generated the training data. Thus, an AE can be used as part of an anomaly detection pipeline\cite{Cozzolino,Liao,Nguyen}: (1) Train an AE on data drawn from a distribution of interest; (2) Feed data  potentially drawn from a different distribution, and measure reconstruction error $\left\|x-f(x)\right\|$ (errors above a threshold are deemed anomalous). By construction, AEs learn salient information that is characteristic of the training data, and hence are robust to the presence of anomalies in the training data \footnote{However, what constitutes an ``anomaly'' in the data may not actually be indicative of an anomalous behavior with respect to cybersecurity. The feature engineering technique we propose does ensure this correlation holds.}. 

The input of an AE must be numeric data, and given that cybersecurity data is largely nominal, we employ \emph{feature engineering} to transform the data into a meaningful representation that is (1) compatible with a neural network architecture and (2) directly relevant to the discovery of anomalous behavior. Rather than employ an ad-hoc feature engineering technique, we adapt word2vec to netflow as others have demonstrated in the literature \cite{log2vec, ramstrom_2019, IP2Vec}.

In essence, word2vec refers to a collection of natural language processing (NLP) techniques, which embeds words from a set of documents into a real vector space, such that semantics are encoded in geometric properties of the vectors. Training these embeddings involves associating words with their contexts (i.e. their neighboring words in the corpus of documents) so that statistical associations between target words and their context words are encoded into the embedding \cite{Bengio2006, NIPS2014_Levy,mikolov2013efficient, mikolov2013distributed}. 

This methodology is easily adaptable to a cybersecurity context, wherein the ``words'' are entities appearing in log data (such as NetFlow or host logs) and the contexts are the other entities appearing in the log. Thus the word2vec paradigm results in a vector space that places entities with similar behaviors nearby (e.g. IPs are clustered together with other IPs and file paths which are loaded in similar processes are neighbors) \cite{IP2Vec}. When only benign records are used to train entity vectors, we are encoding the data based on what constitutes benign cyber activities on the network.

\subsection{Skip-Gram with Negative Sampling}
\begin{figure}
    \centering
    \includegraphics[scale=0.45]{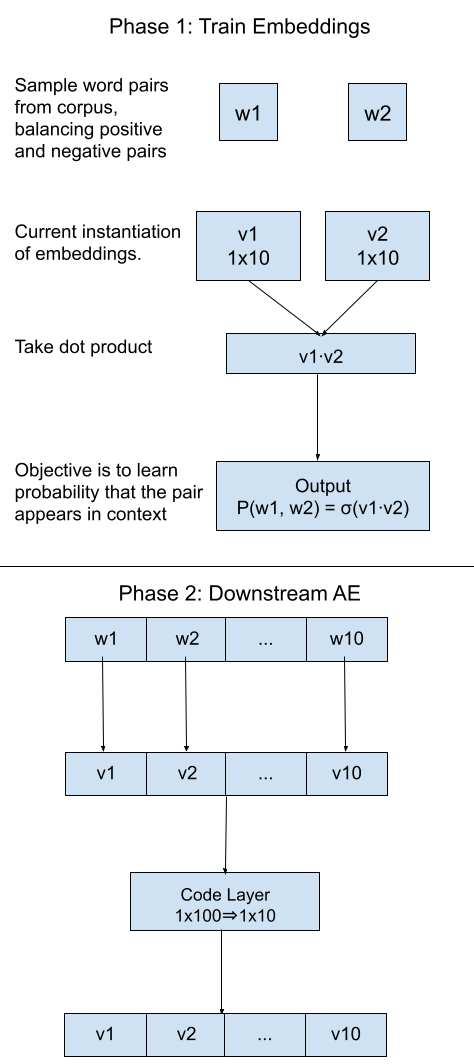}
    \caption{The SGNS method iteratively updates the $10$-dimensional vector embeddings $v_1$ and $v_2$ of the words to optimize BCE loss which forces $\sigma(v_{1}\cdot v_{2})$ to be near $1$ for a positive pair and near $0$ otherwise. Here, $\sigma$ is the sigmoid function. 
    The embeddings corresponding to the entities of a netflow record are concatenated (to form a 100-dimensional vector), which constitutes the input data for an AE.}
    \label{fig:architecture}
\end{figure}

We adapt the Skip-Gram with Negative Sampling (SGNS) variant of word2vec, whose goal is to learn to discriminate between pairs of words that have been observed in the same context (positive pairs), and word pairs that have not (negative pairs) \cite{mikolov2013distributed}. Given a pool of positive and negative pairs, term vector components (i.e. the embedding layer of the model) are updated via gradient descent to optimize binary cross entropy (BCE) loss (with user generated positive or negative labels as ground truth). See Figure \ref{fig:architecture} for more detail. In this work, positive pairs are sampled (with some acceptance probability) from pairs of entities that appear in the same netflow record. Negative pairs are generated by randomly selecting an entity among all entities in the training set to replace the context word observed in the positive pair.

\subsection{Dual Pipeline with UNSW-NB15 Data Set}
We employ the UNSW-NB15 data set to test our implementation of the SGNS methodology. This labeled, synthetic data set consists of enriched netflows containing examples of normal network traffic, as well as examples of cyber attacks from nine different attack categories, including exploits and fuzzers \cite{UNSW1, UNSW2, UNSW3, UNSW4}. While the data contain features from bro and argus, we consider only 10 fields: source/destination IPs, source/destination ports, type of service, durations, and  source/destination byte and packet counts. To reduce the complexity of the model, we only consider those netflows which utilized the TCP protocol (which constitutes roughly 1.5M of the 2.5M total netflows in the entire UNSW corpus).

The pipeline is composed of two parts: a word2vec embedder and an AE anomaly detector, which is similar to that of Ramstrom \cite{ramstrom_2019}. These components are not trained altogether in end-to-end fashion. Word vectors are trained, and then used to construct input to train the AE; reconstruction errors are not back-propagated to adjust the word vector representation. In contrast to the implementation of Ramstrom, which feeds numeric features from the flows directly to the AE, we discretize numeric fields (into quantiles) and embed them, thereby treating them on equal footing with categorical features.

In the first (word2vec) phase, the vocabulary consists of all items found among the positive pairs sampled from the TCP training corpus. In one experiment the positive pairs come only from benign flows, while in another we poison the training set with malicious flows. The negative to positive pair ratio is 7:1, in order to ensure sufficient information is propagated to the embeddings \cite{mikolov2013distributed}.

To represent a netflow record, we concatenate the SGNS trained embeddings corresponding to the entities in the record, and use this as the input to an AE with a single hidden (code) layer. We train an AE using mean squared error loss (MSE). A hold out set is then fed through the trained AE and reconstruction errors for each are measured. When deployed, records with unusually high reconstruction error would then be reported as suspicious.

\subsection{Experimental Paradigm}\label{sec:exp}
 We propose a simple set of experiments: (1) train the pipeline (word vectors and autoencoder) on only benign NetFlows; (2) train the pipeline from scratch with a sufficient number of records corresponding to the exploits class of attacks so that roughly 0.75\% of the training data is malicious; (3) a re-run of (2), but with roughly 1.5\% of the training data coming from malicious flows. We compare distributions of reconstruction errors on benign flows and two classes of malicious traffic: exploits and fuzzers. We also report true positive (malicious) and true negative (benign) rates as well as their harmonic mean, which we call the ``F1'', as a function of MSE cutoff. This will demonstrate the effect of the poisoning on the ability of the AE to detect anomalies. Poisoning levels are based on the widely accepted notion that malicious activity is rare (below 1\%) in the records as compared to normal activities within an enterprise network, a testament to strong firewalls.

\section{Results}\label{sec:results}
\begin{figure}
    \centering
    \includegraphics[width=\columnwidth]{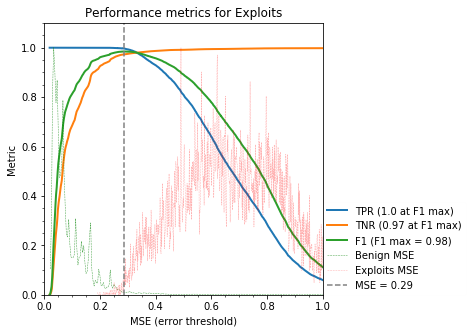}
    \includegraphics[width=\columnwidth]{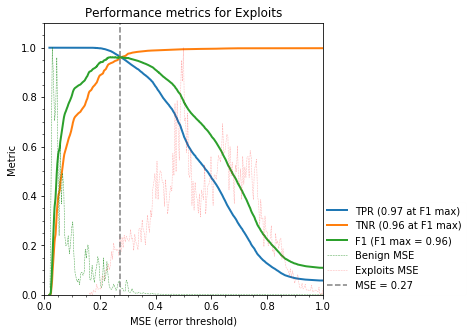}
    \includegraphics[width=\columnwidth]{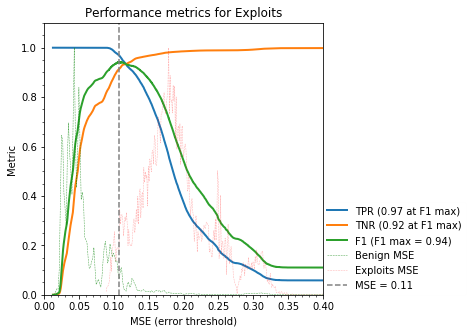}
    \caption{Top: With no poisoning the AE easily distinguishes benign traffic from exploits.
    Middle: With poisoning at 0.75\%, the AE demonstrates a slight decrease in performance, mostly in a decreasing TPR at the optimal F1.
    Bottom: When poison is doubled performance degrades further, due to benign records in the tail over the optimal threshold. Notice the support of the malicious distribution has shifted to the left as compared to the other scenarios.}
    \label{fig:fuzzers}
\end{figure}

\begin{figure}
    \centering
     \includegraphics[width=\columnwidth]{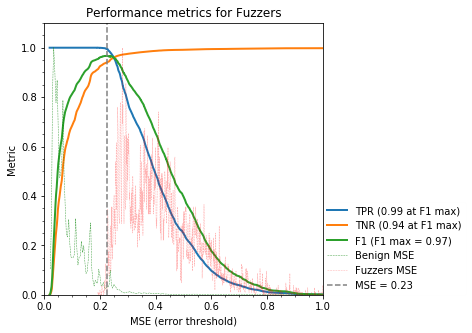}
    \includegraphics[width=\columnwidth]{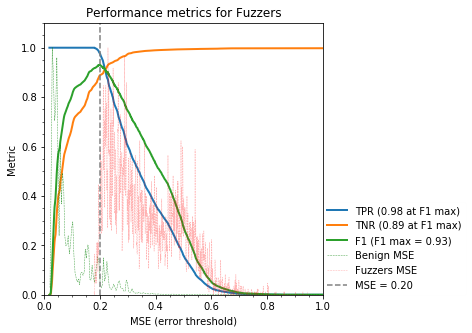}
    \includegraphics[width=\columnwidth]{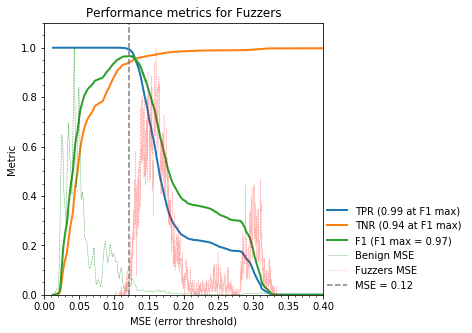}
    \caption{Top: With no poisoning the AE is easily distinguishes benign traffic from fuzzers.
    Middle: With poisoning at 0.75\% with exploits, performance degrades but this is explained by the skewness of the fuzzers distribution, which hardly changes.
    Bottom: When poison is doubled performance actually improves further, likely due to stochasticity of the model. Again the support of the malicious distribution has shifted to the left as compared to the other scenarios.}
    \label{fig:exploits}
\end{figure}

\indent \indent We first present results for an AE trained solely on benign traffic. Figure \ref{fig:exploits} and Figure \ref{fig:fuzzers} show performance metrics for detecting malicious traffic in the test set from the exploits and fuzzers attack classes (resp.). As seen in the figures, the pipeline is largely able to separate benign traffic from both kinds of malicious traffic. As expected, poisoning the training set with exploits examples leads to greater impacts on model performance for exploits records than they do for fuzzers records, and vice-versa. The results show the AE demonstrates a high degree of resilience even when presented with uncharacteristically high levels of poison.

\section{Discussion}\label{sec:discussion} 





\indent \indent The sizes of some effects we observed were small, and their statistical significance has not yet been assessed. The effects of increased poisoning should also be investigated. At small levels of poisoning, it is possible that random factors may obscure relationships we would like to observe. In addition, the results we have shown are derived from single train and test runs. Therefore, they should be considered to be preliminary because of a number of sources of variation. These include random selection of records included in the training set, positive pair selection (for word vector training), selection of negative context words in negative sampling, as well as randomness involved in word vector (embedding layer) and autoencoder training. We plan to address these by sampling in future work.

In order to increase the fraction of exploits records to 1.5\% of the poisoned training set, it was necessary to reduce the number of benign records in the training set. This resulted in a sizeable reduction in the number of destination ports included in the training set. This could be related to the compressed range of observed MSE for this set compared to MSE distributions for unpoisoned and 0.75\% poisoned data sets.

The figures demonstrate that poisoning with a single attack type can affect model performance on different attack types in different ways. This seems reasonable, since the the joint distribution of features need not be the same for all attack classes. However, one should also consider the importance of diversity that may be present within a single attack class. If an attack class is heterogeneous, it is possible for poisoned training to improve performance on the attack class used for poisoning. This counter-intuitive phenomenon might occur if a small sample from an attack class is used to poison a model. Model parameters could adjust to better fit the instances in that sample, while worsening the fit to the other members of class. In other words, a small sample used to poison the training data may not adequately represent all sub-populations present in a single attack class. However, if one wishes to observe such an effect, sampling employed to dampen other sources of variability must be carefully designed.

\bibliography{bib.bib}
\bibliographystyle{plain}
\end{document}